\documentclass[aps,prd,superscriptaddress,showpacs,secnumarabic, eqsecnum]{revtex4}
\usepackage{amsmath,amssymb}

\usepackage{graphicx}

\newcommand{\be}{\begin{eqnarray}}
\newcommand{\ee}{\end{eqnarray}}

\newcommand{\pp}{\partial}

\begin{document}

\title{A new type of scalar field inflation}


\author{Cristiano Germani}
\affiliation{ LUTH, Observatoire de Paris, CNRS UMR 8102, Universit\'{e} Paris Diderot,
5 Place Jules Janssen, 92195 Meudon Cedex, France}
\author{Alex Kehagias}
\affiliation{ Physics Division, National Technical University of Athens,
15780 Athens,  Greece}

\begin{abstract}
In this short note we discuss the possibility of producing a slow rolling inflationary background by considering a massive scalar field whose kinetic term is non-minimally coupled to gravity.
\end{abstract}

\maketitle


\section{Introduction}
The latest cosmological data \cite{wmap}  agree impressively with the assumption that our Universe is, at large scales, homogeneous, isotropic and spatially flat, {\it i.e.}, that it is well described by a Friedmann-Robertson-Walker (FRW) spatially flat geometry. This observation is however a theoretical puzzle. A flat FRW Universe is in fact an extremely fine tuned solution of Einstein equations with normal matter \cite{dodelson}. In the last twenty years or so many attempts have been put forward to solve this puzzle (see for example \cite{many}-\cite{bounce}).
The most developed and yet simple idea still remain inflation. Inflation solves the homogeneity, isotropy and flatness problems in one go just by postulating a
rapid expansion of the early time Universe post Big Bang. However, a fundamental realization of this idea is still eluding us. Originally,
the effective theory of inflation has been realized by sourcing General Relativity (GR) with a slow ``rolling'' massive scalar field \cite{chaotic} with canonical or even
non-canonical kinetic term \cite{k}, whereas, the non-minimal coupling to gravity of a scalar field kinetic term was explored in \cite{pnflation} (see also \cite{mota} for a slightly different realization of the same idea). There, it was showed that a non minimally coupled three-form to gravity, able to reproduce a cosmic inflationary background, is dual to an inflating 0-form (scalar field) whose kinetic term is, indeed, non minimally coupled to gravity.

In this note, inspired by the three-form case of \cite{pnflation}, we will discuss about the possibility of obtaining a slow rolling inflationary background with a scalar field whose kinetic term is non-minimally coupled to gravity (for special cases see \cite{amendola}).

Previous attempts to reproduce a (fast rolling) quasi-DeSitter background with a massless scalar fields whose kinetic terms are non-minimally coupled to gravity can be found in \cite{others}.

\section{Inflation}

Standard chaotic inflation is obtained by using the following scalar field action minimally couple to gravity\footnote{Note that the Newtonian coupling $2\kappa^2$ has been re-absorbed in $\Phi$ for simplicity.}
\be
S=\int d^4x \sqrt{-g} \left[R-\frac{1}{2}g^{\alpha\beta}\partial_{\alpha}\Phi\partial_\beta\Phi-\frac{1}{2}m^2\Phi^2\right] \ ,\label{action}
\ee
where $m$ is the scalar field mass.

On a FRW background
\be
ds^2=-dt^2+a^2 dx\cdot dx\ ,
\ee
the scalar field equation, {\it i.e.} variation of (\ref{action}) with respect to $\Phi$, turns out to be
\be
\ddot \Phi+3H\dot\Phi+m^2\Phi=0\ ,\label{scalar}
\ee
where the notation $\dot f=df/dt$ and $H=\dot a/a$ has been used.

The only independent Einstein equation obtained from the variation of (\ref{action}) with respect to $g^{\alpha\beta}$ is
\be
H^2=\frac{1}{12}\dot\Phi^2+\frac{1}{12}m^2\Phi^2\ .\label{ei}
\ee
An inflationary background is an almost DeSitter expansion ($H\simeq const.$) of the Universe. This is easily obtained whenever the slow roll conditions 
\be\label{consistency}
\dot \Phi^2\ll m^2\Phi^2 \ , \ddot \Phi\ll 3H\dot\Phi\ ,
\ee
are satisfied. In this case, and during slow roll, the scalar field evolves as
\be\label{slow}
\dot\Phi\simeq -\frac{m^2\Phi}{3H}\ ,
\ee
whereas the time variation of the spacetime curvature remains almost constant with respect to the Hubble expansion, {\it i.e.}
\be\label{desitter}
\Big|\frac{\dot R}{(6H)^3}\Big|\simeq \frac{m^2}{9H^2}\ll 1\ .
\ee
The last condition is a consistency condition obtained by taking the time variation of (\ref{slow}) and comparing it with the slow roll conditions (\ref{consistency}). By the symmetries of the background geometry, the condition (\ref{desitter}) approximately defines a DeSitter cosmology.

\section{A new scalar field action}

Let us now turn our attention into possible modifications of (\ref{action}). The most general non-minimally coupled scalar field action to gravity can be recast into the following form
\be
S=\int d^4x \sqrt{-g} \left[\left(1+\alpha U(\Phi)\right)R-\frac{1}{2}\Lambda^{\alpha\beta}\partial_{\alpha}\Phi\partial_\beta\Phi-V(\Phi)\right] \ ,\label{action2}
\ee
where $U(\Phi)$ is a non-minimally coupled potential with coupling constant $\alpha$ and $\Lambda^{\alpha\beta}$ is a tensor obtained from curvatures contractions. We will restrict our analysis by setting $\alpha=0$ and $V(\Phi)=1/2m^2\Phi^2$, although it can been easily generalized. In this case, the action (\ref{action2}) reduces to
\be
S=\int d^4x \sqrt{-g} \left[R-\frac{1}{2}\Lambda^{\alpha\beta}\partial_{\alpha}\Phi\partial_\beta\Phi-\frac{1}{2}m^2\Phi^2\right]\label{action3}\ .
\ee
If $\Lambda^{\alpha\beta}$ is not trivial ({\it i.e.} it is not only proportional to the metric) then this action generically contains higher derivatives (in the gravity equations) arising from the variation with respect to $g^{\alpha\beta}$. Although this seems to directly bring ghost instabilities, in some cases this does not happen \cite{perturbations}. We will however avoid to discuss further this issue here and postpone it for a future work. 

An approximate DeSitter solution of the theory (\ref{action3}) during slow roll, might be obtained by an appropriate restriction of the tensor $\Lambda^{\alpha\beta}$ which we shall discuss in the following. The key point to achieve this goal is to keep the slow rolling properties of the field $\Phi$.

Let us then start by analyzing the scalar field equations.

Suppose an almost DeSitter solution, {\it i.e.} $R_{\alpha\beta}\simeq \mbox{const.}\times g_{\alpha\beta}$, during a slow rolling phase of the field $\Phi$, exists. Thus, the combination $L_{\alpha\beta}\equiv R_{\alpha\beta}-\frac{1}{4} g_{\alpha\beta} R\simeq 0$. Very schematically, one then might consider a form for $\Lambda^{\alpha\beta}$ to be the inverse of an analytic tensorial function $\Delta_{\alpha\beta}=\Delta_{\alpha\beta}(g_{\alpha\beta},L_{\alpha\beta})$. The tensor $\Delta_{\alpha\beta}$ can therefore be expanded, during slow roll, as follows
\be\label{Delta}
\Delta_{\alpha\beta}=g_{\alpha\beta}+a_1L_{\alpha\beta}+a_2 L_{\mu\beta}L^\mu_{\alpha}+\ldots\ ,
\ee
where $a_i$ are dimensionfull coefficients such that, schematically, $a_i (L_{\alpha\beta})^i\ll g_{\alpha\beta}$ during slow roll. In this case then, if the modified Einstein equations coming from (\ref{action3}) admit an approximate DeSitter background, the scalar equations (\ref{slow}), together with the slow roll conditions (\ref{consistency}), are approximately reproduced. 

We now discuss the gravity equations by assuming again that an almost DeSitter solution exist during slow roll. The consistency of this assumption restricts the form of $\Lambda^{\alpha\beta}$, as we shall show in the following.

To obtain the Einstein equations one can use the property
\be
\Lambda^{\mu\alpha}\Delta_{\mu\beta}= \delta^{\alpha}_\beta\ .\label{id}
\ee
The first order variation of (\ref{id}) is then
\be
\delta \Lambda^{\alpha\beta}=-\Lambda^{\mu\alpha}\Lambda^{\nu\beta}\delta \Delta_{\mu\nu}\ .
\ee
At lowest order in slow roll one therefore has
\be
\delta \Lambda^{\alpha\beta}\simeq-\delta \Delta_{\mu\nu} g^{\mu\alpha}g^{\nu\beta}\ ,
\ee
or, explicitly,
\be
\delta \Lambda^{\alpha\beta}\simeq-(\delta g_{\mu\nu}+a_1\delta L_{\mu\nu})g^{\mu\alpha}g^{\nu\beta}\ ,
\ee
as all the higher orders $L^i$ are zero at the lowest slow roll order.

The consistency conditions for the existence of an almost DeSitter background can now be checked.

As it was said before, a sufficient conditions for which the background spacetime can be considered an almost DeSitter manifold is that the time derivative of the Ricci curvature is small compared to the Hubble expansion, {\it i.e.}
\be
\Big|\frac{\dot R}{(6H)^3}\Big|=\Big|\frac{\dot T}{(6H)^3}\Big|\ll{\cal O}(1)\ .\label{cond1}
\ee
In (\ref{cond1}) $T$ is the trace of the energy momentum tensor related to $\Phi$ and the first equality comes from the Einstein equations obtained by varying (\ref{action3}) with respect to $g^{\mu\nu}$.

After a long but straightforward calculation, at lowest order in slow roll, one finds that the trace of the energy momentum tensor related to the scalar field $\Phi$ is\footnote{Note that $a_1H^2\dot\Phi^2$ is not necessarily of slow roll order.}
\be
T\simeq 9H^2 a_1\dot\Phi^2-m^2\Phi^2\ .
\ee
By using (\ref{slow}), one then finds
\be
 \Big|\frac{\dot T}{(6H)^3}\Big|\simeq \left(a_1m^2+1\right) \frac{m^2}{9H^2}\ll{\cal O}(1)\ .
 \ee
Finally, by using the consistency condition $m^2\ll 9H^2$ coming from a the time derivative of (\ref{slow}), one finds that a sufficient condition to have a DeSitter background out of the scalar field action (\ref{action3}) is
\be
-{\cal O}(1)\frac{1}{m^2}\leq a_1\leq {\cal O}(1)\frac{1}{m^2}\ .
\ee
On a FRW background, one might also consider the field redefinition $\Phi=\frac{1}{ma^3}\partial_t (a^3\phi)$. With this, it turn out that the choices $a_1=-2/m^2$ and $\Delta_{\alpha\beta}=g_{\alpha\beta}+a_1 L_{\alpha\beta}$, exactly (at all order in slow roll), reproduce the field equations (\ref{scalar},\ref{ei}) for the field $\phi$ \cite{pnflation}. In fact, by using these choices, one reproduces the dual scalar field theory of the 3-form inflation of \cite{pnflation}.
In this case however, only small field models (for example $V(\Phi)=V_0+\frac{1}{2}m^2\Phi^2$, where $V_0$ is a constant \footnote{The same analysis developed before applies here.}) are gravitationally stable \cite{grav}. Moreover, as scalar perturbations in this particular model quickly decay at super-horizon scales \cite{perturbations}, they are not able to re-produce the observed temperature fluctuations of the Cosmic Microwave Background (CMB) \cite{wmap}.

\section{Dual three-form action}

Here we prove that the general theory described above is dual to a three-form
theory. To show that, one notes that the
action (\ref{action3}) can be derived by integrating out the auxiliary field $B^\alpha$ in 
the following theory
\be
S=\int d^4 x \sqrt{-g} \left(R-\frac{1}{2}m^2\Phi^2-\mu
B^\alpha\partial_\alpha\Phi+
\frac{\mu^2}{2}\Delta^{\alpha\beta}
B_{\alpha} B_{\beta}\right)\ . \label{dual3}
\ee
Viceversa, one may integrate out $\Phi$ to obtain an equation for the auxiliary field
$B^\alpha$. After such an integration we get
\be
\Phi=\frac{\mu}{m^2}\pp_\alpha B^\alpha\ .
\ee
Thus, assuming that $\Phi$ is the dual scalar of the four-form 
\be
F_{\mu\nu\rho\sigma}=m \epsilon_{\mu\nu\rho\sigma} \Phi\ ,
\ee
where $\epsilon_{\mu\nu\rho\sigma}$ is the volume element and  
\be
F_{\mu\nu\rho\sigma}=\nabla_\mu A_{\nu\rho\sigma}-\nabla_\sigma
A_{\mu\nu\rho}+\nabla_\rho 
A_{\sigma\mu\nu}-\nabla_\nu A_{\rho\sigma\mu}\ ,
\ee
we find that 
\be
A_{\mu\nu\rho}=\frac{\mu}{m} \epsilon_{\mu\nu\rho\alpha}B^\alpha\ .
\ee
The dual action obtained after integrating out $\Phi$ in (\ref{dual3}) may be
therefore written as the following three-form theory
\be
S_3=\int d^4 x \sqrt{-g} \left(
R-\frac{1}{48}F_{\mu\nu\rho\sigma}F^{\mu\nu\rho\sigma}-\frac{1}{2}A_{\mu\nu\kappa}\,
M^{\kappa\lambda}\,{A_\lambda}^{\mu\nu}\right) \label{action323}\ ,
\ee
where 
\be
M_{\mu\nu}=\frac{m^2}{6}\left(\Delta\,  g_{\mu\nu}-3 \Delta_{\mu\nu}\right)\, ,
\ee
and $\Delta=g^{\mu\nu} \Delta_{\mu\nu}$.
 
With $\Delta_{\mu\nu}$ given by (\ref{Delta}), the action (\ref{action323}) may be explicitly written, during slow roll, as
\be
S_3=\int d^4 x \sqrt{-g}\left(
R-\frac{1}{48}F_{\mu\nu\rho\sigma}F^{\mu\nu\rho\sigma}
-\frac{1}{12} m^2 A_{\mu\nu\kappa}A^{\mu\nu\kappa}+\frac{3}{12}a_1 m^2 \left[R^{\kappa\lambda}-\frac{1}{4}R
g^{\kappa\lambda}\right]\,A_{\mu\nu\kappa}
\,{A_\lambda}^{\mu\nu}+\ldots \right)\ ,
\ee
where the dots stands for higher curvature terms.

The value $a_1=-2/m^2$ and $a_i=0$ with $i>2$, corresponds to the 3-form inflation discussed in \cite{pnflation}. We see that in this case the three-form action does not contain higher derivatives. This automatically guarantees that no ghost instabilities related to higher time derivatives in the Einstein equations of the dual scalar theory (\ref{action3}) arise.

\section{Conclusions}

In this note it was discussed about the possibility of introducing non-minimal interactions of a scalar field kinetic term to gravity. By restricting the non-minimal couplings to be of the form of tensors constructed from combination of curvatures such that they would vanish in a DeSitter background, we found the conditions for which the slow rolling properties of a scalar field $\Phi$, driven by a flat enough potential $V$, are not spoiled. In this case, an almost DeSitter (inflating) cosmology is re-produced during the slow rolling phase of $\Phi$. 

Although these new type of scalar field theories can generate an inflationary background, they however deeply differ form the chaotic inflationary case ({\it i.e.} minimally coupled scalar field) of \cite{chaotic} at the perturbative level. For example, for a specific choice of parameters such that the scalar field action is dual to the three form inflation of \cite{pnflation}, the scalar perturbations produced during slow roll decay at super-horizon scales \cite{perturbations}. This is in contrast to the chaotic inflationary case where the (super-horizon) perturbations remain frozen producing the observed spectrum of scalar perturbations in the CMB \cite{wmap}. A three-form inflation must then be assisted (for example by a curvaton \cite{curvaton}) to re-produce the observed scalar perturbations of the CMB. 

To conclude, we would like to stress that the new inflationary models introduced in this note might produce new, observationally testable, signatures in the CMB such as new forms of scalar and tensorial linear perturbations and/or non-gaussianities related to the tensorial structure of the new non-minimal couplings. This important analysis is however out of the scope of this paper and left for future work. 

\acknowledgments
CG thanks Jean-Michel Alimi for organizing a stimulating conference for which this note has been prepared. This work is partially supported by the European Research and Training Network MRTPN-CT-2006 035863-1 and the PEVE-NTUA-2009 program.


\begin{thebibliography}{9}
\bibitem{wmap} 
E.~Komatsu {\it et al.}  [WMAP Collaboration],
  arXiv:0803.0547 [astro-ph].
\bibitem{dodelson}
S.~Dodelson,
{\it  Amsterdam, Netherlands: Academic Pr. (2003) 440 p}

\bibitem{many}
G.~Veneziano,
  Phys.\ Lett.\  B {\bf 265}, 287 (1991).
\bibitem{ekpyrotic}
J.~Khoury, B.~A.~Ovrut, P.~J.~Steinhardt and N.~Turok,
  Phys.\ Rev.\  D {\bf 64}, 123522 (2001)
  [arXiv:hep-th/0103239];  E.~I.~Buchbinder, J.~Khoury and B.~A.~Ovrut,
  Phys.\ Rev.\  D {\bf 76}, 123503 (2007)
  [arXiv:hep-th/0702154].

\bibitem{cyclic}
P.~J.~Steinhardt and N.~Turok,
  Phys.\ Rev.\  D {\bf 65}, 126003 (2002)
  [arXiv:hep-th/0111098].
\bibitem{gas}
A.~Nayeri, R.~H.~Brandenberger and C.~Vafa,
  Phys.\ Rev.\ Lett.\  {\bf 97}, 021302 (2006)
  [arXiv:hep-th/0511140].
\bibitem{slingshot}
C.~Germani, N.~E.~Grandi and A.~Kehagias,
  Class.\ Quant.\ Grav.\  {\bf 25}, 135004 (2008)
  [arXiv:hep-th/0611246]; arXiv:0706.0023 [hep-th];  C.~Germani and M.~Liguori,
  Gen.\ Rel.\ Grav.\  {\bf 41}, 191 (2009)
  [arXiv:0706.0025 [astro-ph]].
  C.~Germani, N.~Grandi and A.~Kehagias,
  arXiv:0706.0023 [hep-th].
C.~Germani, N.~Grandi and A.~Kehagias,
  AIP Conf.\ Proc.\  {\bf 1031} (2008) 172
  [arXiv:0805.2073 [hep-th]].

\bibitem{bounce}  P.~Peter, E.~J.~C.~Pinho and N.~Pinto-Neto,
  Phys.\ Rev.\  D {\bf 75}, 023516 (2007)
  [arXiv:hep-th/0610205]; M.~Novello and S.~E.~P.~Bergliaffa,
  Phys.\ Rept.\  {\bf 463}, 127 (2008)
  [arXiv:0802.1634 [astro-ph]].
\bibitem{chaotic} A.~D.~Linde,
  Phys.\ Lett.\  B {\bf 129} (1983) 177.
\bibitem{k}  C.~Armendariz-Picon, T.~Damour and V.~F.~Mukhanov,
  Phys.\ Lett.\  B {\bf 458}, 209 (1999)
  [arXiv:hep-th/9904075].

 \bibitem{pnflation}
 C.~Germani and A.~Kehagias,
  JCAP {\bf 0903} (2009) 028
  [arXiv:0902.3667 [astro-ph.CO]].
 \bibitem{mota}
T.~S.~Koivisto, D.~F.~Mota and C.~Pitrou,
  JHEP {\bf 0909} (2009) 092
  [arXiv:0903.4158 [astro-ph.CO]].
  \bibitem{amendola}
  L.~Amendola,
  Phys.\ Lett.\  B {\bf 301} (1993) 175
  [arXiv:gr-qc/9302010].
  \bibitem{others}
  S.~Capozziello, G.~Lambiase and H.~J.~Schmidt,
  Annalen Phys.\  {\bf 9} (2000) 39
  [arXiv:gr-qc/9906051];
   S.~Capozziello and G.~Lambiase,
  Gen.\ Rel.\ Grav.\  {\bf 31} (1999) 1005
  [arXiv:gr-qc/9901051];
  S.~F.~Daniel and R.~R.~Caldwell,
  Class.\ Quant.\ Grav.\  {\bf 24} (2007) 5573
  [arXiv:0709.0009 [gr-qc]];
S.~V.~Sushkov,
  Phys.\ Rev.\  D {\bf 80} (2009) 103505
  [arXiv:0910.0980 [gr-qc]].
\bibitem{grav}
T.~Kobayashi and S.~Yokoyama,
  JCAP {\bf 0905} (2009) 004
  [arXiv:0903.2769 [astro-ph.CO]].
  
\bibitem{perturbations}
C.~Germani and A.~Kehagias,
  JCAP {\bf 0911} (2009) 005
  [arXiv:0908.0001 [astro-ph.CO]].

\bibitem{curvaton}
D.~H.~Lyth and D.~Wands,
  Phys.\ Lett.\  B {\bf 524} (2002) 5
  [arXiv:hep-ph/0110002].

\end{thebibliography}
\end{document}